\documentclass[%
 a4paper,
 reprint,
 amsmath,amssymb,
 aps
]{revtex4-1}

\usepackage[utf8]{inputenc}
\usepackage{amsmath,amsfonts,amstext,amssymb,mathrsfs}
\usepackage{graphicx}

\usepackage{hyperref}
\usepackage{xcolor}

\hypersetup{
    colorlinks,
    linkcolor={blue!50!black},
    citecolor={blue!50!black},
    urlcolor={blue!80!black}
}

\newcommand{\I}{\mathrm{i}}
\newcommand{\D}{\mathrm{d}}

\newcommand{\R}{\mathbb{R}}

\newcommand{\<}{\langle}
\renewcommand{\>}{\rangle}

\newcommand{\nn}{\nonumber}

\newcommand{\kk}{\bar{\rho}^2_h}
\newcommand{\da}{\Delta A}
\newcommand{\db}{\Delta B}

\newcommand{\red}[1]{{\color{red}{#1}}}
\newcommand{\blue}[1]{{\color{blue}{#1}}}
\newcommand{\green}[1]{{\color{green!50!black}{#1}}}

\newcommand\mat[4]{{\left( \begin{array}{cc} {#1} & {#2} \\ {#3} & {#4} \end{array} \right)}}

\definecolor{LightGray}{rgb}{0.85,0.85,0.85}

\begin{document}

\title{Wormholes from two-sided $T \bar{T}$-deformation}

\author{Adam Bzowski}
\email{adam.bzowski@physics.uu.se}
\affiliation{Department of Physics and Astronomy, Uppsala University, Box 516, SE-75120, Uppsala, Sweden}

\date{\today}

\begin{abstract}

\noindent We introduce a new coupling between stress tensors of the CFTs living on the two boundaries of the BTZ black hole. Similar to the $T \bar{T}$-deformation, the system exhibits universal properties and is solvable. The resulting geometry is an extreme case of a wormhole with the right and left BTZ wedges glued together along the horizons. We show that the geometry is realized by uniform shock waves emanating from both asymptotic boundaries. The construction has profound implications for the structure of the Hilbert space of states of the dual QFT.

\end{abstract}

\maketitle

\section{Introduction}

In their seminal work \cite{Gao:2016bin} Gao, Jafferis and Wall have shown that a wormhole can be opened by coupling the two CFTs living on the two boundaries of the BTZ black hole. With the interaction turned on only for a brief moment, the coupling sends a shock wave through the bulk, which opens a wormhole, \cite{Shenker:2013pqa,Shenker:2013yza,Roberts:2014isa,Stanford:2014jda}. 

While the GJW wormhole has been extensively studied, \textit{e.g.}, \cite{vanBreukelen:2017dul,Bak:2018txn,Haehl:2019fjz,Hirano:2019ugo,Freivogel:2019whb,Fu:2018oaq,Freivogel:2019lej}, a possible caveat is the use of perturbative analysis. On numerous occasions it has been pointed out, \cite{Fu:2018oaq,Freivogel:2019lej,Adams:2006sv,Maldacena:2018lmt,Horowitz:2019hgb,Arefeva:2019ugp}, that non-perturbative, non-semi-classical corrections are crucial in the construction of wormholes.

The aim of this paper is to modify the GJW construction in order to present the full, non-perturbative analysis of a 3-dimensional wormhole. By coupling the \emph{stress tensors} of the two boundary theories one constructs a universal and completely solvable set-up. The deformation resembles the famous $T \bar{T}$-deformation, \cite{Smirnov:2016lqw,Cavaglia:2016oda}, but in our case it is a \emph{2-sided} deformation.

We will show how the 2-sided $T \bar{T}$-deformation changes the topology of the manifold, leading to the opening of an extreme case of a wormhole. The result is confirmed by the construction of the geometry of \emph{uniform shocks}: a spacetime filled with shock waves emitted uniformly from both boundaries. While the problem of multiple shocks was tackled before, \cite{Shenker:2013yza,Roberts:2014isa,Hirano:2019ugo}, the metric of uniform shocks filling the spacetime was not constructed.

We show that -- while all semi-classical observables remain identical to the BTZ observables -- the structure of the Hilbert space of states changes. The degrees of freedom on both boundaries are not independent; they are reduced by `half' with respect to the BTZ black hole. The situation is similar to the proposals of \cite{tHooft:1984kcu,Domenech:1987gw,Hooft:2016itl,Hooft:2016vug,Betzios:2016yaq,Chen:2019ror,Balasubramanian:2020hfs,Strauss:2020rpb,Bzowski:2018aiq}.

\section{Two-sided $T \bar{T}$-deformation}

\subsection{Definition} \label{sec:def}

\begin{figure}[htb]
	\includegraphics[width=0.50\textwidth]{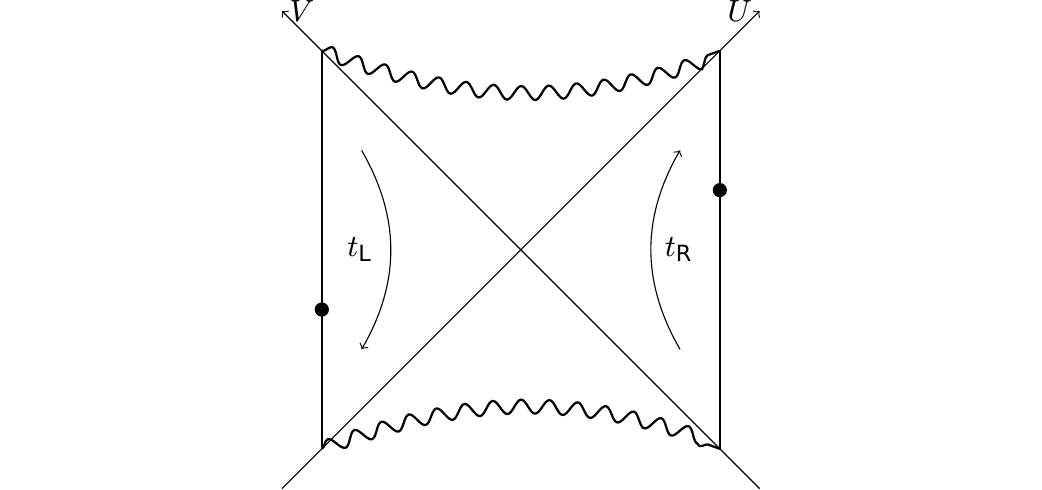}
	\centering
	\caption{The Penrose diagram of the BTZ black hole. The two dots represent the corresponding points with $t = t_L = t_R$. \label{fig:btz}}
\end{figure}

We start with the eternal BTZ black hole, \cite{Banados:1992wn}, whose Schwarzschild metric in the left and right wedges is
\begin{align} \label{btz:rho}
\D s^2 & = - (\rho_{I}^2 - \rho_h^2) \D t_{I}^2 + \frac{l^2 \D \rho_{I}^2}{\rho_{I}^2 - \rho_h^2} + \rho_{I}^2 \D \phi_{I}^2,
\end{align}
where $I = L, R$, $l$ denotes the AdS length, $\rho_h$ -- the Schwarzschild radius, $t_{L,R} \in \R$ and $\phi_{L,R}$ are the coordinates on a circle of radius $1$. The Penrose diagram is presented in Figure \ref{fig:btz}.

Motivated by \cite{Gao:2016bin}, we want to couple the QFTs living on the two asymptotic boundaries by coupling their stress tensors, $T_{\mu\nu}^L$ and $T_{\mu\nu}^R$. First, we must decide at which points the two operators are evaluated. To make things simple, with time flowing in `opposite' directions in the two wedges, we identify $t = t_R = t_L$ and $\phi = \phi_R = \phi_L$, as shown in Figure \ref{fig:btz}.

In \cite{Gao:2016bin} the two boundary operators were scalars. Here we have the additional issue of how to contract the indices of the two stress tensors. Since turning on the source will generically break Lorentz invariance, we can follow the case of non-Lorentz-invariant $T \bar{T}$-deformation in \cite{Cardy:2018jho} and define the interaction
\begin{align} \label{Sint}
S_{int} = \sum_{\mu\nu\rho\sigma} \lambda_{\mu\nu\rho\sigma} \int \D t \D \phi \, T_{\mu\nu}^L(t, \phi) T_{\rho\sigma}^R(t, \phi)
\end{align}
for a set of couplings $\lambda_{\mu\nu\rho\sigma}$. We will simplify the discussion by considering only four non-vanishing couplings,
\begin{align} \label{int}
& S_{int} = \int \D t \D \phi \, \left[ \lambda_{tttt} T_{tt}^L T_{tt}^R + \lambda_{\phi\phi\phi\phi} T_{\phi\phi}^L T_{\phi\phi}^R \right.\nn\\
& \qquad \left. + \: 4 \lambda_{t \phi t \phi} T_{t \phi}^L T_{t \phi}^R + \lambda_{t t \phi \phi} ( T_{tt}^L T_{\phi\phi}^R + T_{\phi\phi}^L T_{tt}^R  ) \right].
\end{align}
A special case corresponds to $T_{\mu\nu}^L T^{\mu\nu R}$, which we define using the Minkowski metric to raise and lower all indices,
\begin{align} \label{TTbar}
T_{\mu\nu}^L T^{\mu\nu R} & = T_{tt}^L T_{tt}^R + T_{\phi\phi}^L T_{\phi\phi}^R - 2 T_{t \phi}^L T_{t \phi}^R.
\end{align}
A particularly interesting case corresponds to the $T \bar{T}$-like deformation. In the standard prescription, \cite{Smirnov:2016lqw,Cavaglia:2016oda}, the $T \bar{T}$ operator equals $T_{zz} T_{\bar{z}\bar{z}} - T_{z \bar{z}}^2$ in the Euclidean setting. We can define its 2-sided generalization as a `symmetrization' of the $T \bar{T}$ operator over the two sides. In Lorentzian signature one finds $T^{(L} \bar{T}^{R)} = \frac{1}{8} T_{\mu\nu}^L T^{\mu\nu R}$, the simple coupling in \eqref{TTbar}.

From the point of view of the QFTs, the deformation is difficult to define. For this reason, we turn to the holographic definition. The idea is that from the point of view of the theory living on the \emph{left} boundary the interaction term \eqref{Sint} introduces a source for the stress tensor proportional to the \emph{right} stress tensor. We look for a \emph{global} bulk metric, which near the left and right boundaries admits the Fefferman-Graham expansion,
\begin{align} \label{zmet}
\D s^2 = \frac{1}{z_{I}^2} \left[ l^2 \D z_{I}^2 + \left( \gamma^{I}_{\mu\nu (0)} + O(z_I^2) \right) \D x^\mu \D x^\nu \right],
\end{align}
with $x = t, \phi$. $\gamma^{L,R}_{\mu\nu (0)}$ are identified as left and right boundary metrics, while $\gamma^{L,R}_{\mu\nu (2)}$ determine the expectation value of the stress tensor, \cite{deHaro:2000vlm},
\begin{equation} \label{Tij}
\< T^{L,R}_{\mu\nu} \>_s = \frac{1}{8 \pi G_N l} \left( \gamma^{L,R}_{\mu\nu (2)} - \gamma_{\mu\nu (0)}^{L,R} \frac{l^2}{2} R^{L,R}_{(0)} \right),
\end{equation}
where $R_{(0)}^{L,R}$ is the Ricci scalar of $\gamma_{\mu\nu (0)}^{L,R}$. We \emph{define} the theory deformed by \eqref{Sint} as the theory satisfying the following constraints,
\begin{align}
& \gamma^{L}_{\mu\nu (0)} - \eta_{\mu\nu} = \sum_{\rho\sigma} \lambda_{\mu\nu\rho\sigma} \< T_{\rho\sigma}^R \>_s, \label{bnd1} \\
& \gamma^{R}_{\mu\nu (0)} - \eta_{\mu\nu} = \sum_{\rho\sigma} \lambda_{\rho\sigma\mu\nu} \< T_{\rho\sigma}^L \>_s. \label{bnd2}
\end{align}
We supplement this system of equations with the initial condition w.r.t. the couplings: when all couplings vanish, the bulk metric should converge to the BTZ metric in the left and right wedges.

Note that from the point of view of a single boundary the interaction \eqref{Sint} behaves like a single trace rather than a double trace deformation. If we dropped the assumption of a single global metric, the system of equations (\ref{bnd1}, \ref{bnd2}) would emerge from a 1-sided deformation by a product of two operators. Such a system was analyzed in \cite{Aharony:2005sh,Betzios:2019rds} in the context of scalar operators.

\subsection{The solution}

To find the deformed geometry we consider the BTZ metric in the Fefferman-Graham coordinates \eqref{btz:z} with the Lorentzian and Euclidean tensors $\eta_{\mu\nu}$ and $\delta_{\mu\nu}$ replaced,
\begin{align} \label{ansatz}
\D s^2 & = \frac{l^2 \D z_I^2}{z_I^2} + \left[ \left( \frac{1}{z_I^2} + \frac{z_I^2}{z_h^4} \right) \tilde{\eta}^I_{\mu\nu} + \frac{2}{z_h^2} \tilde{\delta}^I_{\mu\nu} \right] \D x^\mu \D x^\nu
\end{align}
where
\begin{align} \label{bnd_eta}
& \tilde{\eta}^{I}_{\mu\nu} = \mat{-A^{I}}{0}{0}{B^{I}}, && \tilde{\delta}^{I}_{\mu\nu} = \mat{A^{I}}{0}{0}{B^{I}}
\end{align}
and $I = L,R$. The constants $A^{L,R}$ and $B^{L,R}$ depend on the couplings in \eqref{int} and the symmetry requires $A = A^L = A^R$ and $B = B^L = B^R$. Furthermore, when all couplings vanish, we should recover the BTZ geometry, which means $A = 1 + O(\lambda_{\mu\nu\rho\sigma})$ and $B = 1 + O(\lambda_{\mu\nu\rho\sigma})$. The ansatz \eqref{ansatz} satisfies Einstein equations and, with the flat boundary, the system (\ref{bnd1}, \ref{bnd2}) simplifies to a pair of algebraic equations. The solution reads
\begin{align}
A & = \frac{1 - \kk (\lambda_{\phi\phi\phi\phi} + \lambda_{tt \phi\phi})}{1 + \kk ( \lambda_{tttt} - \lambda_{\phi\phi\phi\phi}) + \bar{\rho}_h^4(\lambda_{tt \phi\phi}^2 - \lambda_{tttt} \lambda_{\phi\phi\phi\phi})}, \label{A} \\
B & = \frac{1 + \kk (\lambda_{tttt} + \lambda_{tt \phi\phi})}{1 + \kk ( \lambda_{tttt} - \lambda_{\phi\phi\phi\phi}) + \bar{\rho}_h^4(\lambda_{tt \phi\phi}^2 - \lambda_{tttt} \lambda_{\phi\phi\phi\phi})}, \label{B}
\end{align}
where $\kk = \rho_h^2/(16 \pi G_N l)$. Notice that as with the standard $T \bar{T}$-deformation, \cite{Smirnov:2016lqw,Cavaglia:2016oda,McGough:2016lol}, the solution becomes singular for some special values of the couplings.

The boundary metric, $\tilde{\eta}_{\mu\nu}$, is not related by a symmetry (Lorentz or rescaling) to the Minkowski metric. This suggests a possibility of a superluminal propagation, \cite{McGough:2016lol,Adams:2006sv}, with the new speed of light $c = \sqrt{A/B}$.

From the bulk point of view the deformed geometry \eqref{ansatz} in the Schwarzschild coordinates reads
\begin{align} \label{def:rho}
\D s^2 & = - A (\rho_I^2 - \rho_h^2) \D t^2 + \frac{l^2 \D \rho_I^2}{\rho_I^2 - \rho_h^2} + B \rho_I^2 \D \phi^2.
\end{align}
With arbitrary $\gamma > 0$, the substitution
\begin{align} \label{to_btz}
& t' = \gamma \sqrt{A} t, && \phi' = \gamma \sqrt{B} \phi, && \rho'_{L,R} = \gamma^{-1} \rho_{L,R},
\end{align}
brings the deformed metric into the BTZ metric, with the new Schwarzschild radius $\rho_h' = \gamma^{-1} \rho_h$. Following the reasoning of \cite{McGough:2016lol,Kraus:2018xrn,Wang:2018jva,Bzowski:2018pcy} one would conclude that the deformed geometry is that of another BTZ black hole and therefore nothing has happened. This is incorrect, at least for the 2-sided deformation, as we will discuss next.

\section{The wormhole}

\begin{figure}[htb]
	\includegraphics[width=0.48\textwidth]{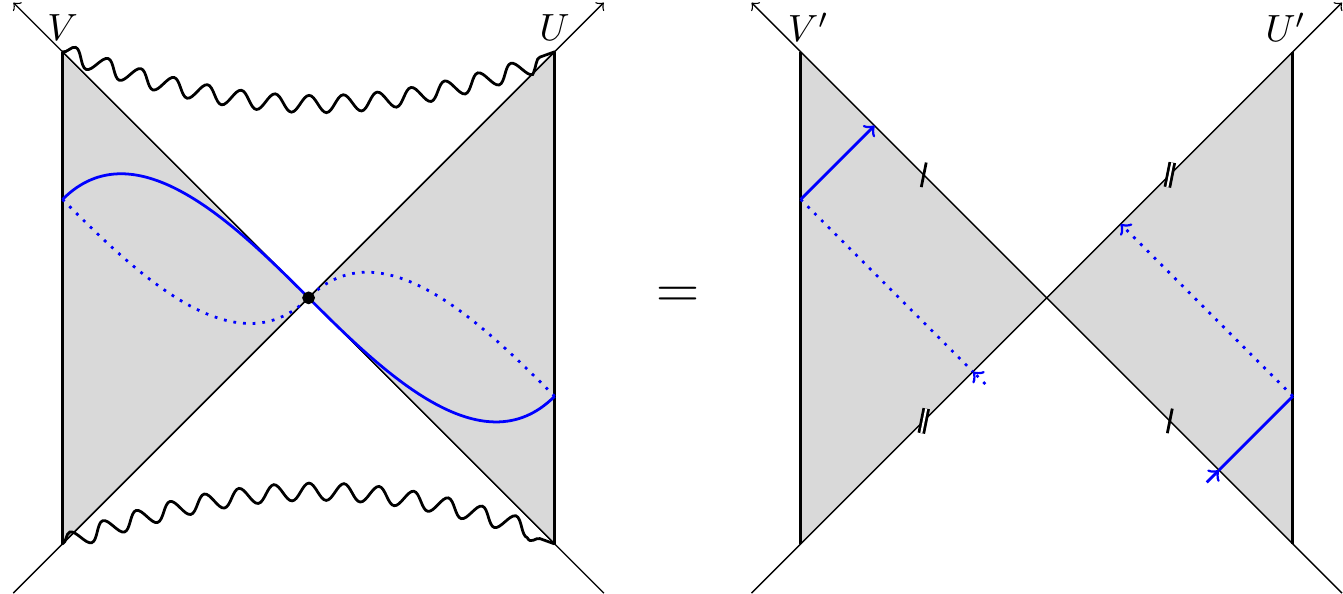}
	\centering
	\caption{Geometry of the wormhole. Left: in the metric \eqref{def:uv} all geodesics pass through $(0,0)$. Right: by using (\ref{to_BTZ_UV_U}, \ref{to_BTZ_UV_V}) this corresponds to the identifications along the horizons: $(U',0) \sim (-U',0)$ and $(0,V') \sim (0,-V')$.\label{fig:wh}}
\end{figure}

\subsection{A problem and a resolution}

The critical issue with the transformation \eqref{to_btz} is that it does not extend to the entire manifold. To be specific, in the Kruskal coordinates, the transformation \eqref{to_btz} takes form
\begin{align}
& | U' | = | U |^{\frac{1}{2} (1 + \sqrt{A})} |V|^{\frac{1}{2} (1 - \sqrt{A})}, \label{to_BTZ_UV_U} \\
& | V' | = | U |^{\frac{1}{2} (1 - \sqrt{A})} |V|^{\frac{1}{2} (1 + \sqrt{A})}, \label{to_BTZ_UV_V}
\end{align}
where the signs of $U'$ and $V'$ are the same as the signs of $U$ and $V$ in the respective wedges. The transformation is valid in all open wedges, but cannot be extended to the entire spacetime. For $A < 1$ the transformation maps the whole variety $U V = 0$ to the single point $U' = V' = 0$, while for $A > 1$ it blows up at $U = V = 0$. Note that this would not have been a problem if the compactification of the wedges had not already been fixed. 

What we propose is that the deformation \eqref{int} altered the topology of the spacetime beyond the horizons. In the left and right wedges the deformed geometry is that of the BTZ black hole in the primed coordinates \eqref{to_btz}, but the relation between the two wedges has changed. As we will argue in the remainder of this section, for $A < 1$, the wormhole has been opened and the two wedges are glued together along the $U' = 0$ and $V' = 0$ horizons. The null line of constant $U'$ in the right wedge is glued to the null line of $-U'$ in the left wedge. Analogously, the lines of constant $V'$ and $-V'$ are glued together. The resulting geometry is presented in Figure \ref{fig:wh}. The geometry exhibits closed timelike loops and thus it can be regarded as an extreme case of a wormhole, where every null ray emitted from one boundary hits the opposite boundary. Note that such a severe alteration of the deep IR is consistent with all our assumptions in Section \ref{sec:def}, since the conditions (\ref{bnd1}, \ref{bnd2}) are only well-defined in their respective wedges. Similar causal structure was discovered in \cite{Adams:2006sv,Canate:2018chw,Canate:2019sbz} in the Einstein-Maxwell theory with higher derivative corrections. The proposal resembles that of the antipodal identification
\cite{tHooft:1984kcu,Domenech:1987gw,Hooft:2016itl,Hooft:2016vug,Betzios:2016yaq,Chen:2019ror,Balasubramanian:2020hfs,Strauss:2020rpb}, although we identify points along the horizons only.

In the following sections we will provide the support for the construction in Figure \ref{fig:wh} by analyzing its geodesic structure and showing that the geometry emerges from multiple shock waves.

\subsection{Geodesics}

\begin{figure*}[htb]
	\includegraphics[width=1.00\textwidth]{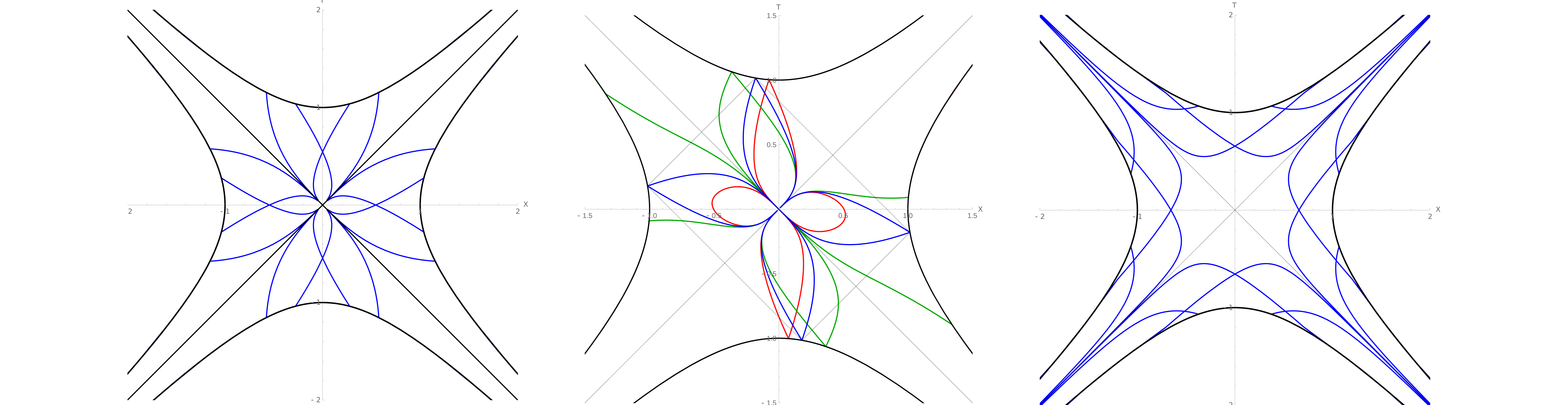}
	\centering
\caption{Examples of zero-angular momentum geodesics in the deformed geometry \eqref{def:uv}. Null geodesics are in \blue{blue}, timelike in \red{red}, and spacelike in \green{green}. Left and center: geodesics in the wormhole geometry, $A < 1$: note that all geodesics pass through $U = V = 0$. Right: geodesics in the geometry with $A > 1$: the geodesics are repelled from $U V = 0$.\label{fig:geo}}
\end{figure*}

In Kruskal coordinates the deformed geometry \eqref{def:rho} becomes
\begin{align} \label{def:uv}
\D s^2 & = \D s_{\text{BTZ}}^2 + \frac{\da \: l^2}{(1 + U V)^2} \left[ 2 \D U \D V - \frac{V}{U} \D U^2 - \frac{U}{V} \D V^2 \right] \nn\\
& \qquad\qquad + \db \: \rho_h^2 \frac{(1 - U V)^2}{(1 + U V)^2} \D \phi^2,
\end{align}
where $\D s^2_{\text{BTZ}}$ is the original BTZ metric \eqref{btz:uv}, while $A = 1 - \da$ and $B = 1 + \db$. As expected, the metric becomes singular at $U V = 0$. Notice that this prevents us from following \cite{Gao:2016bin} and using the ANEC criterion, \cite{Wall:2009wi}, as a test for a wormhole. It also casts doubt on the applicability of the perturbative analysis there.

The geodesics can be obtained simply by substituting the transformations (\ref{to_BTZ_UV_U}, \ref{to_BTZ_UV_V}) to the geodesic equation for the BTZ black hole, \eqref{geo}. It is easy to see that for $\da < 0$ all geodesics pass through $U = V = 0$, while for $\da > 0$ they all diverge to $U \rightarrow \infty$ when $V \rightarrow 0$. Therefore, apart from $U = V = 0$, the variety $U V = 0$ cannot be probed, which suggests that the geodesics may continue directly from one wedge to another, as shown in Figure \ref{fig:wh}. The situation is similar to the wormhole constructed in \cite{Maeda:2008bh}. Examples of various geodesics are plotted in Figure \ref{fig:geo}.

\subsection{Uniform shock waves}

The coupling of the boundary theories for a short period of time introduces a shock wave into the bulk and a wormhole can form, \cite{Shenker:2013pqa,Shenker:2013yza,Roberts:2014isa,Stanford:2014jda}. The resulting geometry has been studied extensively, \cite{Dray:1984ha,Hotta:1992qy,Sfetsos:1994xa}, and is characterized by a non-vanishing $UU$ (or $VV$) component of the metric localized on the shock wave. This results in a sudden kick when a particle crosses the shock wave. In the BTZ background the shift in the null coordinate equals,
\begin{align} \label{DeltaU}
\D U = E e^{\frac{t \rho_h}{l}} \delta(V) \D V,
\end{align}
where $t$ denotes the boundary time of the shock wave emission.

We want to derive the geometry of \emph{uniform shocks}, where the shock waves are emitted from both boundaries with uniform energy. To this end we use \eqref{DeltaU} as the defining property, which must be satisfied by null geodesics. To smear it uniformly in time, we use the fact that near the boundary $\delta(V) \sim \frac{l}{\rho_h} e^{t \rho_h/l} \delta(t)$ and thus
\begin{align} \label{shock_geo}
\D U = \frac{E l}{\rho_h} e^{\frac{2 t \rho_h}{l}} \D V = \frac{E l}{\rho_h} \left( - \frac{U}{V} \right) \D V,
\end{align}
where we used \eqref{to_btz:uv}. Alternatively, the extra factor of $e^{t \rho_h/l}$ can be explained as in \cite{Shenker:2013yza}: it is the relative energy of two shock waves separated by $V$.

It is easy to check that the null geodesic obtained by equating \eqref{to_BTZ_UV_U} to a constant satisfies \eqref{shock_geo}, while the symmetric geodesic obtained from \eqref{to_BTZ_UV_V} satisfies the symmetric version of \eqref{shock_geo} with the same value of $E$, with $U$ and $V$ exchanged. This proves that the deformed geometry \eqref{def:uv} is the geometry of uniform shocks. Furthermore, it shows that both wedges contain shock waves traveling in both left and right directions. Since the shocks must emanate from the boundaries, the right-moving shock waves in the right wedge can only originate from the left boundary (and \textit{vice versa}). This supports the global structure presented in Figure \ref{fig:wh}.

\subsection{Hilbert space structure}

Here we describe how the Hilbert space structure of the boundary theory has been altered. Consider a scalar field quantized on the background in Figure \ref{fig:wh}. All semi-classical observables, which do not depend on non-perturbative, beyond-the-horizon effects, must be identical to the BTZ case, including the the fact that the vacuum is the thermofield double state, \cite{Maldacena:2001kr}. (The fact that the vacuum state of the wormhole must be very similar to the thermofield double was argued in \cite{Maldacena:2018lmt,Maldacena:2019ufo}.) What changes is the structure of the Hilbert space itself. While in the case of the BTZ black hole the Hilbert space is the tensor product $\mathcal{H} = \mathcal{H}_L \otimes \mathcal{H}_R$ of the Hilbert spaces of the left and right QFTs, this is no longer true for the wormhole. The subsequent analysis should take place in the primed coordinates, but for notational simplicity, we will drop all primes.

To carry out the canonical quantization in the wormhole background one solves the Klein-Gordon equation. In Schwarzschild coordinates one finds two mode functions, $f_{\omega k}(\rho_I)$, $I = L,R$. Normalizable modes exhibit the specific fall-off at the boundary, $f_{\omega k}(\rho_I) \sim \rho_I^{-\Delta}$, where $\Delta > 1$ satisfies $m^2 l^2 = \Delta(\Delta - 2)$. See Appendix \ref{app:btz} for details.

If one starts with the mode $f_{\omega k}(\rho_R)$ in the right wedge, its continuation to the left wedge can be obtained from the expression \eqref{hor_exp}. Since the future-directed geodesic in the right wedge becomes past-directed in the left wedge, as seen in Figure \ref{fig:wh}, positive frequency modes in one wedge continue to negative frequency modes in the other wedge. This leads to
\begin{align}
f_{\omega k}(\rho_L) = e^{\frac{\beta \omega}{2}} \tilde{f}^{\ast}_{\omega, -k}(\rho_L),
\end{align}
where tilde denotes the analytic continuation of the mode from the right wedge to the left and $\beta = 2 \pi l / \rho_h$. This implies the operator relation between the left and right creation-annihilation operators,
\begin{align} \label{a_rel}
& a_{\omega, -k}^{L} = e^{-\frac{\beta \omega}{2}} a_{\omega, k}^{R \dagger}, && a_{\omega, -k}^{R} = e^{-\frac{\beta \omega}{2}} a_{\omega, k}^{L \dagger}.
\end{align}
The excitations on both boundaries are not independent. The Hilbert space $\mathcal{H}$ of the dual theory is not the tensor product. Instead, it is isomorphic to the Hilbert space of a single side. The isomorphism, however, does not preserve the Fock space structure. Indeed, it is impossible to uphold both $a^L_{\omega, k} | 0 \> = 0$ and $a^R_{\omega, k} | 0 \> = 0$ in $\mathcal{H}$.

Instead, $\mathcal{H}$ can be constructed as a subspace of the tensor product $\mathcal{H}_L \otimes \mathcal{H}_R$. States in $\mathcal{H}$ are defined as those that satisfy relations \eqref{a_rel}. These are precisely the relations satisfied by mirror operators in the Papadodimas-Raju proposal of state-dependence, \cite{Papadodimas:2013wnh,Papadodimas:2013jku,Papadodimas:2015jra,Papadodimas:2015xma}. In particular, the thermofield double
satisfies these relations.

\section{Summary}

In the paper we have shown how the 2-sided $T \bar{T}$-like deformation of the BTZ black hole accounts for the opening of a wormhole. The global geometry, presented in Figure \ref{fig:wh}, is that of uniform shocks: a geometry obtained from the emission of shock waves uniformly from both boundaries. From the point of view of the boundary theory the BTZ geometry and the wormhole geometry differ by non-perturbative, beyond-the-horizon, invisible to semi-classical physics, effects. This results in the severe change in topology as well as in the structure of the Hilbert space of states.

\begin{acknowledgments}

I would like to thank Marjorie Schillo for the encouragement and help in the preparation of this manuscript. I am supported by the Knut and Alice Wallenberg Foundation under Grant No.~113410212.
 
\end{acknowledgments} 

\appendix

\section{BTZ geometry} \label{app:btz}

The Schwarzschild form of the BTZ metric is given in \eqref{btz:rho}. It can be brought to the Fefferman-Graham form,
\begin{align} \label{btz:z}
\D s^2 & = \frac{l^2 \D z_I^2}{z_I^2} + \left[ \left( \frac{1}{z_I^2} + \frac{z_I^2}{z_h^4} \right) \eta_{\mu\nu} + \frac{2}{z_h^2} \delta_{\mu\nu} \right] \D x_I^\mu \D x_I^\nu
\end{align}
where $I = L,R$, $z_h = 2/\rho_h$, $x_I = t_I, \phi_I$ and the conformal boundaries are located at $z_I = 0$.

The Kruskal coordinates are defined by the following substitutions
\begin{align} \label{to_btz:uv}
& \exp \left( \frac{2 \rho_h t_I}{l} \right) = - \frac{U}{V}, && \frac{\rho_I}{\rho_h} = \frac{1 - U V}{1 + U V}
\end{align}
and the metric becomes
\begin{align} \label{btz:uv}
\D s^2 & = \frac{4 l^2}{(1 + U V)^2} \left[ - \D U \D V + \frac{\rho_h^2}{4 l^2} (1 - U V)^2 \D \phi^2 \right].
\end{align}
The right wedge corresponds to $U > 0$ and $V < 0$ and its future horizon is located at $V = 0$. The left wedge corresponds to $U < 0$ and $V > 0$. The conformal boundaries are at $U V = - 1$, while the singularities at $U V = 1$. The Penrose diagram is presented in Figure \ref{fig:btz}.

In Kruskal coordinates zero-angular momentum geodesics satisfy
\begin{align} \label{geo}
(C - U)(1 + C V) = \kappa V,
\end{align}
where $C$ and $\kappa$ are integration constants. The sign of $\kappa$ determines the type of the geodesic: $\kappa < 0$ is timelike, $\kappa > 0$ spacelike, and $\kappa = 0$ is null.

The solution to the Klein-Gordon equation $(\Box - m^2) \Phi = 0$ in the BTZ background in the Schwarzschild coordinates \eqref{btz:rho} reads
\begin{align}
& \Phi^{(\pm)}_{\omega k}(t, \rho, \phi) = \frac{c_{\omega k}^{(\pm)}}{\sqrt{2 \omega}} e^{-\I \omega t + \I k \phi} \left( \frac{\rho^2}{\rho^2_h} \right)^{a} \left( \frac{\rho^2_h}{\rho^2 - \rho_h^2} \right)^{a + \frac{\Delta_{\pm}}{2}} \times \nn\\
& \qquad \times {}_2 F_1 \left( \frac{\Delta_{\pm}}{2} + a - b, \frac{\Delta_{\pm}}{2} + a + b; \Delta_{\pm}; \frac{\rho_h^2}{\rho_h^2 - \rho^2} \right),
\end{align}
where we dropped the $L,R$ indices, $a = \frac{\I k l}{2 \rho_h}$ and $b = \frac{\I \omega l}{2 \rho_h}$ while $\Delta_{\pm}$ are two roots of the equation $m^2 l^2 = \Delta (\Delta - 2)$. Solutions $\Phi^{(\pm)}_{\omega k}$ behave like $\rho^{-\Delta_{\pm}}$ as $\rho \rightarrow \infty$ while near the horizons in the right wedge,
\begin{align} \label{hor_exp}
\Phi_{\omega k}^{(\pm)} \sim \frac{1}{\sqrt{2 \omega}} \left[ U^{-2b} \beta_{\omega k}^{(\pm)} + (-V)^{2b} \beta_{\omega, -k}^{(\pm) \ast} \right],
\end{align}
where $\beta_{\omega k}^{(\pm)}$ satisfy $\beta_{\omega, -k}^{(\pm)} = \beta_{\omega k}^{(\pm)}$ and $\beta_{\omega k}^{(\pm) \ast} = \beta_{-\omega, k}^{(\pm)}$. Exact expressions for the constants $c_{\omega k}^{(\pm)}, \beta_{\omega k}^{(\pm)}$ and further details can be found in \cite{Papadodimas:2012aq}.

\end{document}